\documentclass[conference]{IEEEtran}

\usepackage[OT1]{fontenc}
\usepackage[utf8]{inputenc}

\usepackage[cmex10]{amsmath}
\usepackage{bm}
\usepackage{amsfonts}
\usepackage{amssymb}
\usepackage{pgfplots}
\pgfplotsset{compat=newest}
\usepgfplotslibrary{groupplots}
\usepackage{tikz}
\usepackage{multirow}
\usetikzlibrary{matrix, positioning, patterns, shapes, arrows}
\usepackage{xcolor}
\definecolor{Paired-2}{RGB}{166,206,227}
\definecolor{Paired-1}{RGB}{31,120,180}
\definecolor{Paired-4}{RGB}{178,223,138}
\definecolor{Paired-3}{RGB}{51,160,44}
\definecolor{Paired-6}{RGB}{251,154,153}
\definecolor{Paired-5}{RGB}{227,26,28}
\definecolor{Paired-8}{RGB}{253,191,111}
\definecolor{Paired-7}{RGB}{255,127,0}
\definecolor{Paired-10}{RGB}{202,178,214}
\definecolor{Paired-9}{RGB}{106,61,154}
\definecolor{Paired-12}{RGB}{255,255,153}
\definecolor{Paired-11}{RGB}{177,89,40}

\DeclareMathOperator{\sign}{sign}

\DeclareMathOperator*{\argmin}{argmin}

\usepackage{subcaption}
\usepackage{booktabs}
\usepackage[tworuled, linesnumbered, vlined]{algorithm2e}
\usepackage{cuted}
\usepackage{soul} 

\begin{document}
\bstctlcite{IEEEexample:BSTcontrol}
\title{Fast Decoding of Multi-Kernel Polar Codes}

\author{\IEEEauthorblockN{Adam Cavatassi, Thibaud Tonnellier and Warren J. Gross}
\IEEEauthorblockA{Department of Electrical and Computer Engineering\\
McGill University, Montr\'eal, Qu\'ebec, Canada\\
Email: adam.cavatassi@mail.mcgill.ca, thibaud.tonnellier@mail.mcgill.ca, warren.gross@mcgill.ca}}


\maketitle

\begin{abstract}
Polar codes are a class of linear error correction codes which provably attain channel capacity with infinite codeword lengths. 
Finite length polar codes have been adopted into the 5th Generation 3GPP standard for New Radio, though their native length is 
limited to powers of 2. Utilizing multiple polarizing matrices increases the length flexibility of polar codes at the expense 
of a more complicated decoding process. Successive
cancellation (SC) is the standard polar decoder and has time complexity $\mathcal{O}(N\log{N})$ due to its sequential nature. 
However, some patterns in the frozen set mirror simple linear codes with low latency decoders, which allows for a significant reduction 
in SC latency by pruning the decoding schedule. Such fast decoding techniques have only previously been used for 
traditional Ar{\i}kan polar codes, causing multi-kernel polar codes to be an impractical length-compatibility technique with no
fast decoders available. We propose fast simplified successive cancellation decoding node patterns, which are compatible with polar codes constructed with 
both the Ar{\i}kan and ternary kernels, and generalization techniques. We outline efficient implementations, made possible
by imposing constraints on ternary node parameters. We show that fast decoding of multi-kernel polar codes has at least 72\%
reduced latency compared with an SC decoder in all cases considered where codeword lengths are (96, 432, 768, 2304).

\end{abstract}

\IEEEpeerreviewmaketitle

\section{Introduction}

Polar codes are class of error correction codes that have been proven to achieve channel capacity at infinite lengths \cite{Arikan2009}.
As such, they have been selected for control channel use in the 5\textsuperscript{th} generation (5G) 3GPP standard for enhanced mobile broadband \cite{3GPP}.
Since polar codes have now entered the scope of commercial use, researching methods to maximize their practicality is crucial.

However, polar codes have limitations that preclude their immediate practicality. The primary decoding technique is known as successive 
cancellation and is a serial algorithm by nature. This shortcoming implies that polar decoders are latency limited and have undesirable
throughput. Further, polar codes of short to medium codeword length have inferior error correction performance when compared with other
state-of-the-art error correction codes. List decoding allows for a considerable improvement in error correction potential, especially when
the polar code is concatenated with a CRC \cite{Tal2015}. Additionally, fast decoding techniques, such as fast simplified successive cancellation, 
have been proposed, which allow for substantial latency reduction \cite{Sarkis2014}. This algorithm identifies simple linear codes that are 
embedded in the intermediate stages of a successive cancellation decoder. Rather than decoding these stages in the typical manner, the simplified
decoders are applied, significantly reducing the number of decoding operations. 

Another drawback associated with polar codes is that their codeword length is limited to powers of two. This property is a result of the recursive Kronecker
product expansion of the $2 \times 2$ polarizing matrix proposed by Ar{\i}kan. In practice, it is desirable to utilize 
error correction codes that can attain any length. Length-matching techniques, such as puncturing \cite{Niu2013} and shortening \cite{Wang2014}, 
have been proposed, but these methods are not ideal due to additional optimization requirements and decoding
complexity that is not directly related to their codeword length. Nonetheless, puncturing and shortening methods have been incorporated into
the polar code scheme of the 5G standard. 

Multi-kernel (MK) polar codes were proposed in order to overcome the lack of length flexibility in conventional polar coding \cite{Benammar2017}.
This method allows incorporating additional polarizing matrices of size larger than two in conjunction with Ar{\i}kan's matrix to build a
polar code with more flexibility in codeword length. Specifically, a ternary $3 \times 3$ matrix was proposed, and so MK polar codes can have lengths
that are powers of two, powers of three, or a product of both. Although additional polarizing matrices of larger sizes have been
proposed \cite{Lin2015}, among them the $3 \times 3$ matrix offers a desirable combination of a sufficiently high polarization exponent and
only a small increase in decoding complexity.

Fast simplified successive cancellation (Fast-SSC) decoders only exist for conventional polar codes, and so we propose an extension to existing fast decoding techniques that makes them
compatible with MK codes. We offer proofs for fast decoding methods that are not trivially applied to MK codes, and we generalize the techniques
to be usable with any construction of MK polar codes that use both the Ar{\i}kan and ternary polarization matrices. Though generalized implementations 
are possible, we demonstrate that efficient applications of these techniques are permitted under certain code construction constraints. We find that 
MK compatible Fast-SSC reduces computation complexity by a minimum of 72 \% in all cases considered in this study.

The remainder of this paper is organized as follows: Section \ref{section:pc} reviews polar code preliminaries, including encoding and decoding, and 
outlines code optimization methods used for MK polar codes. Section \ref{section:fssc} reviews the existing fast decoding scheme and proposes extensions for 
ternary compatibility along with the necessary proofs. Section \ref{section:analysis} outlines the latency reduction of the fast MK decoder and draws 
comparisons with equivalent length matching techniques.

\section{Polar Codes}
\label{section:pc}

Polar codes, denoted $\mathcal{PC}(N, K)$, are linear block codes that have a codeword length of $N$, 
message length $K$, and rate $R = \frac{K}{N}$. Channel polarization is a phenomenon in which $N$ copies of channel $W$ are transformed 
into $N$ synthetic channels with either increased or decreased reliability relative to $W$ \cite{Arikan2009}. 
The $K$ most reliable channels are designated as the information set $\mathcal{I}$, and selected to transmit
information. The $N-K$ remaining channels comprise frozen set $\mathcal{F}$. 

A message $\bm{a}$ of length $K$ is expanded into the sourceword $\bm{u} = (u_0, u_1, \hdots, u_{N-1})$ by placing the 
elements of $\bm{a}$ into indices from $\mathcal{I}$, while all indices in $\mathcal{F}$ are set to 0 and considered frozen.
The codeword $\bm{x} = (x_0, x_1, \hdots, x_{N-1})$ can be encoded by computing $\bm{x} = \bm{u \cdot G}$, where 
the generator matrix $\bm{G} = \bm{T_2}^{\otimes n}$. In other words, $\bm{G}$ is equal to the Kronecker 
product of the matrix $\bm{T_2} = \bigl[\begin{smallmatrix} 1 & 0  \\ 1 & 1 \end{smallmatrix}\bigr]$ that has been carried out 
$n \in \mathbb{N}^{+}$ times. $\bm{T_2}$ is the polarizing matrix proposed by Ar{\i}kan in \cite{Arikan2009} and will 
be referred to in this paper as the Ar{\i}kan kernel. 

\subsection{Multi-Kernel Polar Codes}
Utilization of alternate polarizing matrices, known as kernels, in the formulation for $\bm{G}$ was proposed in \cite{Gabry2016} 
and outlined the possibility of obtaining polar codes that are not constrained to lengths that are powers of $2$. 
The ternary kernel $\bm{T_3} = \Bigl[\begin{smallmatrix} 1 & 1 & 1 \\ 1 & 0 & 1 \\ 0 & 1 & 1 \end{smallmatrix}\Bigr]$ was proposed as the 
$3 \times 3$ polarizing matrix. $\bm{T_3}$ was shown to be optimal for polarization in \cite{Benammar2017}, though it has a 
polarization exponent that is less than that of $\bm{T_2}$. $\bm{T_3}$ can be used as a Kronecker product constituent in conjunction
with the Ar{\i}kan kernel to produce any polar code of length $N = 2^n3^m$ where $n,m\in \mathbb{N}$. The native 
length flexibility of polar codes is thus improved. However it should be noted that the order of the kernels in the
Kronecker product affects $\bm{G}$. For example, a polar code of length $N = 6$ could be encoded with either $T_2 \otimes T_3$
or $T_3 \otimes T_2$, which are two unique generator matrices. For clarity, we can refer to a kernel vector $k$ that stores the 
sizes of the kernels in order as they pertain to the generator matrix. The generator matrix can then be defined as 
\begin{equation}
	\bm{G} = \bigotimes_{i=0}^{m+n} \bm{T_{k_i}}.
\end{equation}
Observe that the kernel order in a MK Tanner graph is reversed from that of the 
the Kronecker product, as can be seen in Fig. \ref{fig:mk_6}. Further, additional kernels of size higher than 3 have
been proposed, although the Ar{\i}kan and ternary kernel are the most common and least complex to use. As such, we will only
investigate MK polar codes derived from these two kernels. Polar codes that use only the Ar{\i}kan kernel will be further referred
to as Ar{\i}kan polar codes. 

\subsection{Successive Cancellation Decoding}

Decoding of MK polar codes can be accomplished using successive cancellation (SC) \cite{Arikan2009}. The encoding Tanner graph is 
restructured into a tree with $M=n+m$ stages, which visualizes the SC algorithm. SC involves a tree search with left-to-right branch priority, where the
leaf nodes in the tree represent the estimated sourceword $\bm{\hat{u}}$. The top of the tree serves as the decoder input, which is the 
received soft data vector $\bm{y}$ in the form of real log likelihood ratios (LLR). 

\begin{figure}[t]
	\centering
	\hspace{-13mm}
	\begin{subfigure}[t]{.30\textwidth}
		\includegraphics[scale = 0.3]{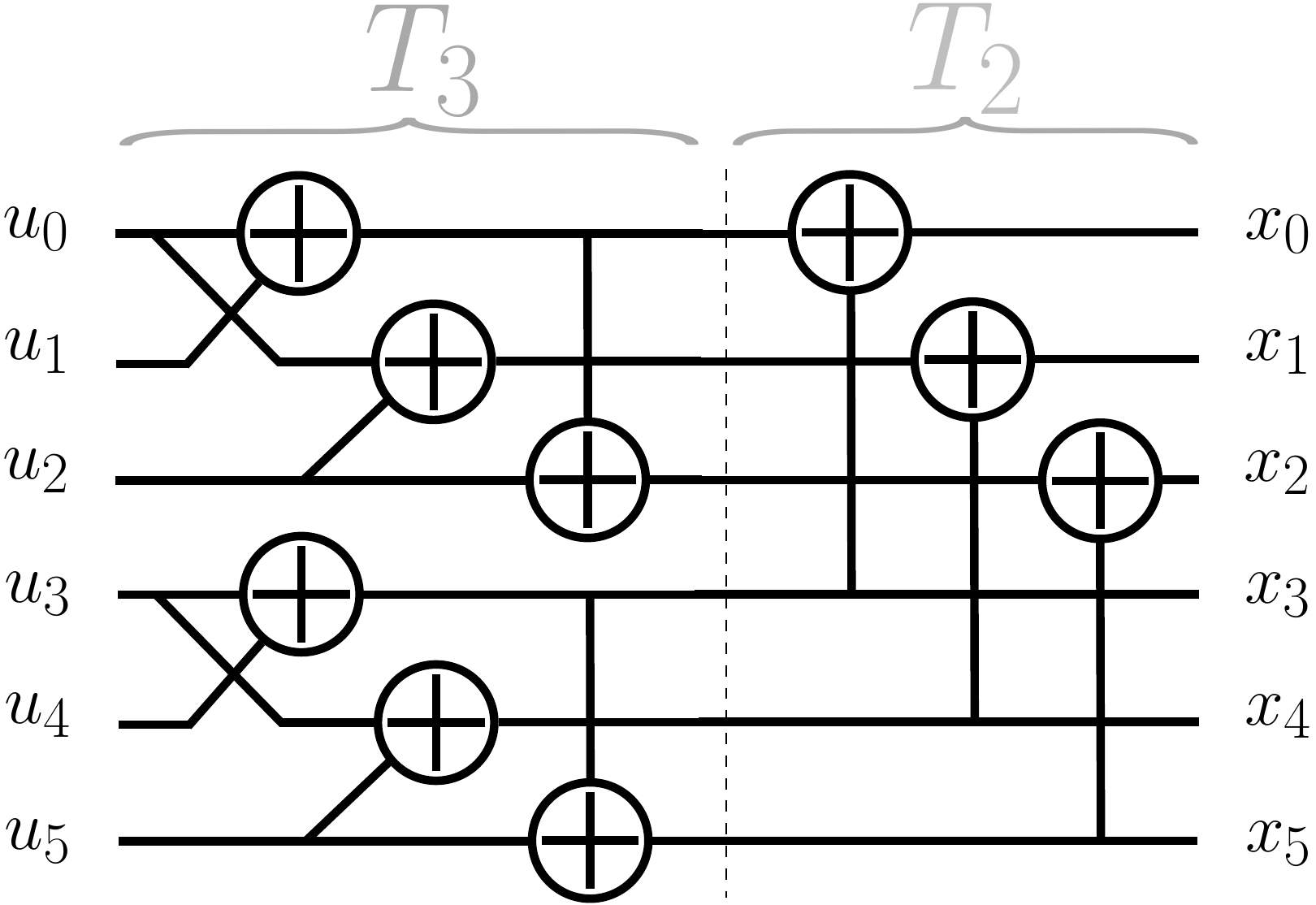}
		\caption{Encoder}
	\end{subfigure}
	\begin{subfigure}[t]{.1\textwidth}
		\includegraphics[scale = 0.3]{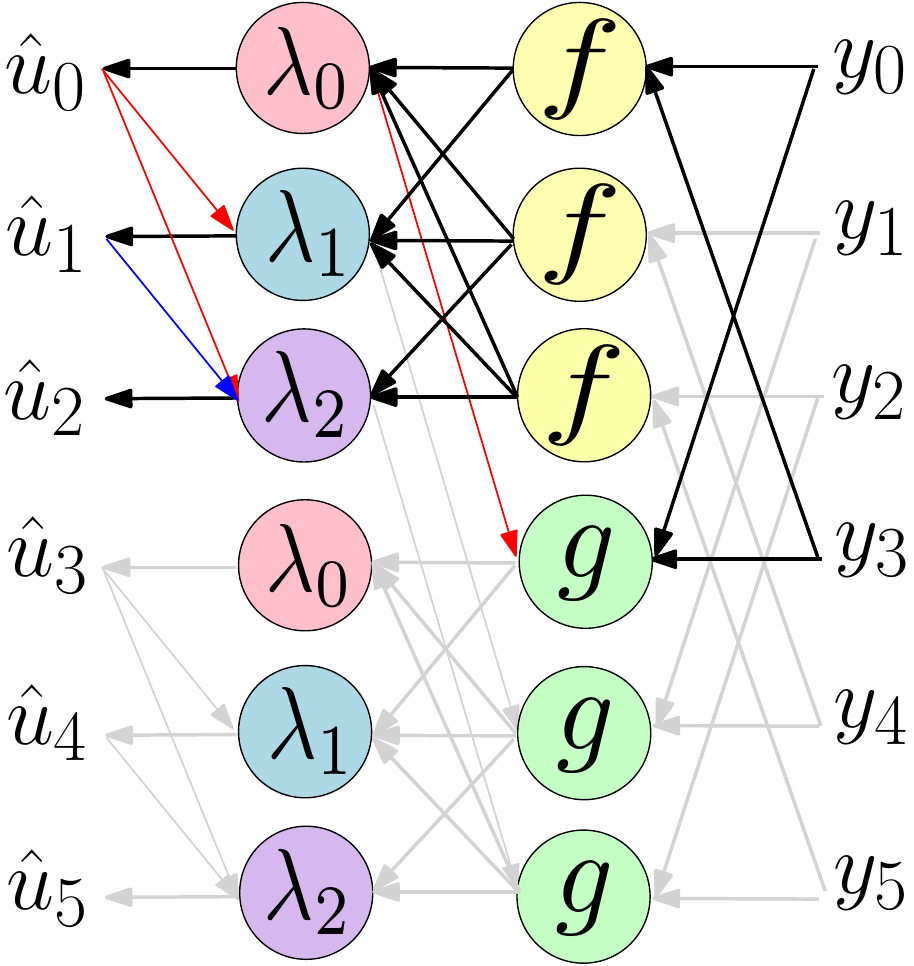}
		\caption{Decoder}
		\label{fig:mk_6:dec}
	\end{subfigure}
	\caption{Multi-kernel polar code with $N=6$ and $G = T_2 \otimes T_3$.}
	\vspace*{-\baselineskip}
	\label{fig:mk_6}
\end{figure}

Beginning from the top of the tree, the branches are traversed by applying LLR transformations and storing the results in the 
proceeding stage. Each stage $S \in \left[0, M\right]$ pertains to either the Ar{\i}kan or ternary kernel and contains $P = \frac{N}{p}$ 
nodes, where $p = \prod_{i=1}^{S-1} k_i$. 
At the bottom stage of the tree, \emph{ie} $S=0$, $p = 1$. Each node in stage $S$ stores both $p$ LLRs and bit partial sums and invokes $\frac{p}{2}$ or 
$\frac{p}{3}$ transformations upon entering, depending on the kernel of stage $S-1$. If the stage $S$ pertains to the Ar{\i}kan kernel, 
the functions $f$ or $g$, found in eq. \ref{eq:arik_dec}, are applied to the left and right branches, respectively. 
\begin{equation}
\label{eq:arik_dec}
	\begin{aligned}
		f(l_0,l_1,) &= l_0 \boxplus l_1, \\
		g(l_0,l_1, u_0) &= (-1)^{u_0} \cdot l_0 + l_1 \\
	\end{aligned}
\end{equation}
In the case that stage $S$ corresponds to the ternary kernel, the functions
$\lambda_0$, $\lambda_1$, and $\lambda_2$ are applied to the left, center, and right branches, respectively.
\begin{equation}
\label{eq:mk_dec}
	\begin{aligned}
		\lambda_0(l_0 ,l_1,l_2) &= l_0 \boxplus l_1 \boxplus l_2, \\
		\lambda_1(l_0 ,l_1,l_2,u_0) &= (-1)^{u_0} \cdot l_0 + l_1 \boxplus l_2, \\
		\lambda_2(l_1,l_2,u_0,u_1) &= (-1)^{u_0} \cdot l_1 + (-1)^{u_0 \oplus u_1} \cdot l_2,
	\end{aligned}
\end{equation}
where $a \boxplus b \approx \sign{(a)}\sign{(b)}\min{(|a|,|b|)}$, $l_0, l_1, l_2$ symbolize LLR values, and 
$u_0, u_1$ denote a partial sum located in a left or center node, respectively. When a leaf node is entered, a hard bit decision is made using the 
LLR stored as
\begin{equation}
\label{eq:hd}
	h_l(l_i) = \begin{cases} 0 & \text{if } l_i > 0 \text{ or } i \in \mathcal{F} \\ 1 & \text{otherwise} \end{cases}.
\end{equation}
A modified version of this hard decision function that neglects whether $i \in \mathcal{F}$ will be referenced throughout this paper as $h()$. 
After computing a bit decision or returning from a right branch, $p$ partial sum updates are executed at the previous node 
before moving down another branch. The partial sum updates are also referred to as \textit{combine} operations. For Ar{\i}kan stages, 
the combine operation is $c_2(s_0, s_1) = (s_0 \oplus s_1, s_1)$, while for a ternary stage it is $c_3(s_0, s_1, s_2) = (s_0 \oplus s_1, s_0 \oplus s_2, s_0 \oplus s_1 \oplus s_2)$.
Ostensibly, an SC decoder can be expressed as schedule of $f$, $g$, $\lambda_0$, $\lambda_1$, and $\lambda_2$ operations, 
where the total number of operations described by $(n+m)N$. Fig. \ref{fig:mk_6:dec} depicts a schedule representation of an SC decoder
that contrasts from the tree representation in Fig. \ref{fig:n18_fast_tree}.

\subsection{Frozen Set Design}

In order to select $K$ indices to constitute $\mathcal{I}$, all $N$ indices must be sorted by reliability. 
Several accurate reliability ordering algorithms exist for conventional polar codes. Among them is Ar{\i}kan's 
Bhattacharyya parameter expansion  \cite{Arikan2009}, which has only been proven to be exact for binary erasure
channels, although it can be used as an approximation when transmitting over a Gaussian channel. A more accurate
reliability representation can be generated using Trifonov's Gaussian Approximation (GA) \cite{Trifonov2012}. The process
assumes all LLRs at each stage of the decoder are Gaussian distributed, and so the absolute value of the mean  
of each index is tracked through each LLR transformation from one stage to the next. To begin, all $N$ indices have the same 
distribution as the channel $W$. Assuming an AWGN channel with mean $0$ and variance $\sigma^2$, each index is initialized as
\begin{equation*}
	 z_i^N = \frac{2}{\sigma^2}=\frac{2}{(2R\frac{E_b}{N_0})^{-1}}=4R\frac{E_b}{N_0} \text{ for } i \in [0,N).
\end{equation*}
To compute the means of the next 
stage, apply the following equations if the stage corresponds to an Ar{\i}kan kernel: 
\begin{align*}
	 z_{2i - 1}^{\frac{N}{2}} &= \phi^{-1}(1-(1-\phi(z_i^N ))^2), \\
	 z_{2i \textcolor{white}{- 0}}^{\frac{N}{2}} &= 2z_i^N. \nonumber
\end{align*}
If the stage corresponds to a ternary kernel, instead apply the following:
\begin{align*}
	z_{3i - 2}^{\frac{N}{3}} &= \phi^{-1}\Big(1\!\!-\!\Big(1\!\!-\!\phi\big(\phi^{-1}(1\!\!-\!(1\!\!-\!\phi(z_i^N ))^2)\big)\Big)(1\!\!-\!\phi(z_i^N))\Big), \nonumber \\
	z_{3i - 1}^{\frac{N}{3}} &= \phi^{-1}(1-(1-\phi(z_i^N ))^2) + z_i^N,  \\
	z_{3i\textcolor{white}{- 0}}^{\frac{N}{3}} &= 2z_i^N, \nonumber
\end{align*}
where $\phi(x)$ and $\phi(x)^{-1}$ are approximated as
\begin{align}
	\phi(x) &= \begin{cases} e^{0.0564 x^2 - 0.485x} & x < 0.8678 \\  e^{\alpha x^\gamma + \beta} & \text{otherwise} \end{cases} \text{ and} \nonumber \\
	\phi^{-1}(x) &= \begin{cases}  4.3049(1 - \sqrt{1 + 0.9567 \log{x}}) & x > 0.6846 \\ ({a \log{x} + b})^{c} & \text{otherwise} \end{cases}, \nonumber
\end{align}
where $\alpha = -0.4527, \beta = 0.0218, \gamma = 0.86, a = \frac{1}{\alpha}, b = \frac{-\beta}{\alpha}$, and $c = \frac{1}{\gamma}$.
These approximations are available from the open source error correction code simulation tool \textit{aff3ct} \cite{Cassagne2017a}. 
The exact formulas devised by Trifinov can be found in \cite{Trifonov2012}.
Proceeding to transform the LLR means for all stages will result in $N$ unique values, which can serve as a basis for which 
to rank the indices in ascending order. From here, $\mathcal{I}$ can be populated with the $K$ highest values and $\mathcal{F}$ will
be composed of the remaining $N-K$ indices. 

\subsection{Kernel Order Optimization}
\label{section:kernel_order}

In \cite{Benammar2017}, it was suggested that the order of kernels for a MK polar code can be optimized by searching all permutations
for the order that produces the highest sum of bit reliabilities among the $K$ best indices. While this method is effective, it is not 
desirable to have this additional optimization step when comparing the practicality of MK polar codes with Ar{\i}kan polar codes. 

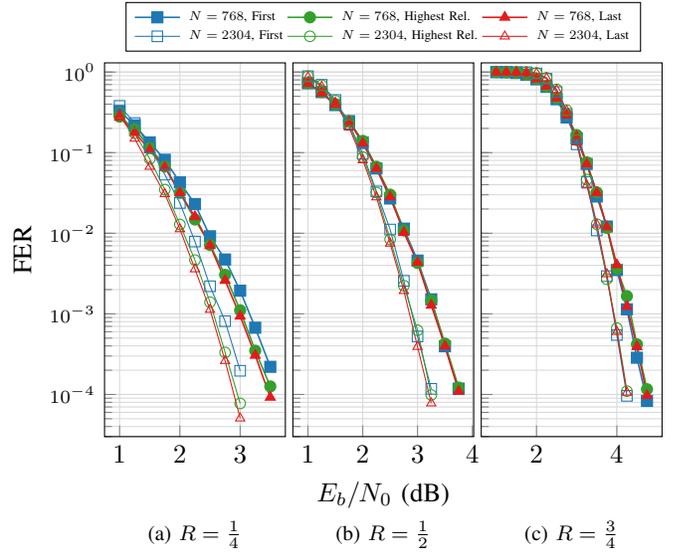
\begin{figure}[t]
	\hspace{5mm}
	\centering
		\begin{tikzpicture}

\begin{groupplot}[
        group style={group name=fer_group, group size= 3 by 1, horizontal sep=1mm,}, height=0.74\linewidth, width=.45\linewidth]

        \nextgroupplot[ylabel=FER, grid=both, grid style={gray!30}, tick align=inside, tickpos=left, yticklabel style = {font=\scriptsize}, ymode=log, ymax=1.3,  ymin=3e-5]
                \addplot[Paired-1, mark=square*, mark size=2pt, semithick] table [x={EbN0}, y={FER}, col sep=comma] {figs/data/768_192/768,192_mk_l.csv};
                \addplot[Paired-3, mark=*, mark size=2pt, semithick] table [x={EbN0}, y={FER}, col sep=comma] {figs/data/768_192/768,192_mk_h.csv};
                \addplot[Paired-5, mark=triangle*, mark size=2pt, semithick] table [x={EbN0}, y={FER}, col sep=comma] {figs/data/768_192/768,192_mk_f.csv};

                \addplot[Paired-1, mark=square, mark size=2pt, thin] table [x={EbN0}, y={FER}, col sep=comma] {figs/data/2304_576/2304,576_mk_l.csv};
                \addplot[Paired-3, mark=o, mark size=2pt, thin] table [x={EbN0}, y={FER}, col sep=comma] {figs/data/2304_576/2304,576_mk_h.csv};
                \addplot[Paired-5, mark=triangle, mark size=2pt, thin] table [x={EbN0}, y={FER}, col sep=comma] {figs/data/2304_576/2304,576_mk_f.csv};

        \nextgroupplot[xlabel={$E_b/N_0$ (dB)}, grid=both, grid style={gray!30}, tick align=inside, tickpos=left, yticklabels={}, ymode=log, ymax=1.3,  ymin=3e-5,
                        legend style={font=\scriptsize, nodes={scale=0.7}, at={(0.5,1.03)}, anchor=south, legend columns=3},
                        legend cell align={left},
                        legend entries={~{$N=768$, First}, ~{$N=768$, Highest Rel.}, ~{$N=768$, Last}, ~{$N=2304$, First}, ~{$N=2304$, Highest Rel.}, ~{$N=2304$, Last}}
                        ]
                \addplot[Paired-1, mark=square*, mark size=2pt, semithick] table [x={EbN0}, y={FER}, col sep=comma] {figs/data/768_384/768,384_mk_l.csv};
                \addplot[Paired-3, mark=*, mark size=2pt, semithick] table [x={EbN0}, y={FER}, col sep=comma] {figs/data/768_384/768,384_mk_h.csv};
                \addplot[Paired-5, mark=triangle*, mark size=2pt, semithick] table [x={EbN0}, y={FER}, col sep=comma] {figs/data/768_384/768,384_mk_f.csv};

                \addplot[Paired-1, mark=square, mark size=2pt, thin] table [x={EbN0}, y={FER}, col sep=comma] {figs/data/2304_1152/2304,1152_mk_l.csv};
                \addplot[Paired-3, mark=o, mark size=2pt, thin] table [x={EbN0}, y={FER}, col sep=comma] {figs/data/2304_1152/2304,1152_mk_h.csv};
                \addplot[Paired-5, mark=triangle, mark size=2pt, thin] table [x={EbN0}, y={FER}, col sep=comma] {figs/data/2304_1152/2304,1152_mk_f.csv};

         \nextgroupplot[grid=both, grid style={gray!30}, tick align=inside, tickpos=left, yticklabels={}, ylabel style = {font=\scriptsize}, ymode=log, ymax=1.3,  ymin=3e-5]
                \addplot[Paired-1, mark=square*, mark size=2pt, semithick] table [x={EbN0}, y={FER}, col sep=comma] {figs/data/768_576/768,576_mk_l.csv};
                \addplot[Paired-3, mark=*, mark size=2pt, semithick] table [x={EbN0}, y={FER}, col sep=comma] {figs/data/768_576/768,576_mk_h.csv};
                \addplot[Paired-5, mark=triangle*, mark size=2pt, semithick] table [x={EbN0}, y={FER}, col sep=comma] {figs/data/768_576/768,576_mk_f.csv};

                \addplot[Paired-1, mark=square, mark size=2pt, thin] table [x={EbN0}, y={FER}, col sep=comma] {figs/data/2304_1728/2304,1728_mk_l.csv};
                \addplot[Paired-3, mark=o, mark size=2pt, thin] table [x={EbN0}, y={FER}, col sep=comma] {figs/data/2304_1728/2304,1728_mk_h.csv};
                \addplot[Paired-5, mark=triangle, mark size=2pt, thin] table [x={EbN0}, y={FER}, col sep=comma] {figs/data/2304_1728/2304,1728_mk_f.csv};

\end{groupplot}
\node[below = 1cm of fer_group c1r1.south] {\footnotesize (a) $R=\frac{1}{4}$};
\node[below = 1cm of fer_group c2r1.south] {\footnotesize (b) $R=\frac{1}{2}$};
\node[below = 1cm of fer_group c3r1.south] {\footnotesize (c) $R=\frac{3}{4}$};

\end{tikzpicture}
	\caption{FER curves for MK polar code with $N=768,2304$ sweeping rates $R=(\frac{1}{4},\frac{1}{2},\frac{3}{4})$ (left to right) comparing kernel ordering strategies.}
	\vspace*{-.5\baselineskip}
	\label{fig:fer:768}
\end{figure}

Fig. \ref{fig:fer:768} demonstrates that for long MK polar codes, the kernel optimization step is largely inconsequential. The figure depicts MK 
polar codes with $N=768$ and $N=2304$ sweeping rates $R=(\frac{1}{4},\frac{1}{2},\frac{3}{4})$ under SC decoding. The codes are constructed using GA for each point 
in the plot to ensure that the frozen sets are optimal throughout the simulation. Three different kernel ordering strategies are investigated. The 
\textit{First} and \textit{Last} labels indicate that kernels are ordered such that the ternary kernels serve as either the first or last components
of the Kronecker product, respectively. \textit{Highest Reliability} ensures that the kernel order is optimized for the highest overall reliability using the method
outlined in the previous paragraph. It can be concluded that placing the ternary kernel in the Kronecker product at either the first or last positions 
produces comparable error correction results against an optimized kernel order for long polar codes. Further, observe that low rate codes have better performance
using \textit{Last}, while the opposite configuration performs best for medium to high rate codes. 


\section{Fast-SSC Decoding}
\label{section:fssc}

The essence of fast SC decoding is to prune the decoding tree to reduce the schedule and thus decrease
latency \cite{Sarkis2014}. This decoding technique is
known as fast simplified SC (Fast-SSC) and works by identifying specific frozen set patterns that mirror embedded subcodes that can be decoded efficiently. These fast nodes are decoded with maximum likelihood, which indicates that FSSC retains the same error correction performance of SC.
This section will outline the four basic fast nodes that are currently known and extend them to be compatible with the $\bm{T_3}$ kernel.

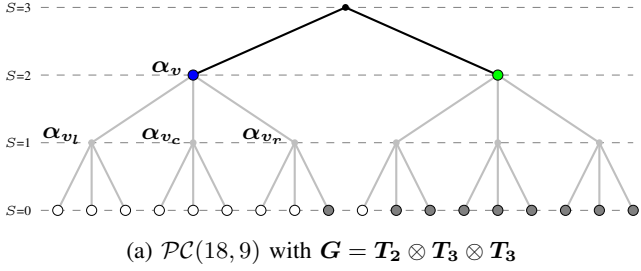
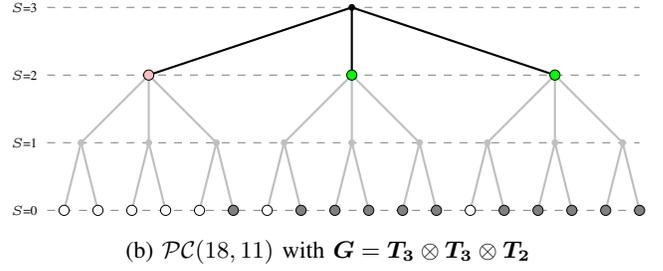
\begin{figure*}[t]
\makebox[\linewidth][c]{
	\centering
	\hspace{0mm}
	\begin{subfigure}[t]{.5\textwidth}
	\centering
			\begin{tikzpicture}[scale=.45, thick]
	
	\node[left] at (.5,0) {\tiny $S$=0};
	\node[left] at (.5,2) {\tiny $S$=1};
	\node[left] at (.5,4) {\tiny $S$=2};
	\node[left] at (.5,6) {\tiny $S$=3};
	
	\draw [very thin,gray,dashed] (.5,0) -- (18,0);
	\draw [very thin,gray,dashed] (.5,2) -- (18,2);
	\draw [very thin,gray,dashed] (.5,4) -- (18,4);
	\draw [very thin,gray,dashed] (.5,6) -- (18,6);

	\foreach \x in {1,4,7,10,13,16}{
	\draw [lightgray] (\x,0) -- (\x+1,2);
	}
	\foreach \x in {2,5,8,11,14,17}{
	\draw [lightgray] (\x,0) -- (\x,2);
	}
	\foreach \x in {3,6,9,12,15,18}{
	\draw [lightgray] (\x,0) -- (\x-1,2);
	}

	\foreach \x in {2,11}{
	\draw [lightgray] (\x,2) -- (\x+3,4);}
	\foreach \x in {5,14}{
	\draw [lightgray] (\x,2) -- (\x,4);}
	\foreach \x in {8,17}{
	\draw [lightgray] (\x,2) -- (\x-3,4);}
	\foreach \x in {5}{
	\draw (\x,4) -- (\x+4.5,6);}
	\foreach \x in {14}{
	\draw (\x,4) -- (\x-4.5,6);}

	\foreach \x in {1,2,3,4,5,6,7,8,10}{
	\fill [white] (\x,0) circle [radius=.15];
	\draw[very thin] (\x,0) circle [radius=.15];	
	}
	\foreach \x in {9,11,12,13,14,15,16,17,18}{
	\fill [gray] (\x,0) circle [radius=.15];
	\draw[very thin] (\x,0) circle [radius=.15];
	}
	
	\foreach \x in {2,5,8,11,14,17}{
	\fill [lightgray] (\x,2) circle [radius=.1];	
	}

	\fill [blue] (5,4) circle [radius=.15];
	\draw[very thin] (5,4) circle [radius=.15];
	\fill [green] (14,4) circle [radius=.15];
	\draw[very thin] (14,4) circle [radius=.15];

	\foreach \x in {9.5}{
	\fill [black] (\x,6) circle [radius=.1];
	}
	
	\node[left] at (5,4.2) {\small $\bm{\alpha_v}$};
	\node[left] at (2,2.2) {\small $\bm{\alpha_{v_l}}$};
	\node[left] at (5,2.2) {\small $\bm{\alpha_{v_c}}$};
	\node[left] at (8,2.2) {\small $\bm{\alpha_{v_r}}$};

	\end{tikzpicture}
		\caption{$\mathcal{PC}(18,9)$ with $\bm{G}=\bm{T_2} \otimes \bm{T_3} \otimes \bm{T_3}$}
		\label{fig:n18_fast_tree:a}
	\end{subfigure}
	\begin{subfigure}[t]{.5\textwidth}
	\centering
			\begin{tikzpicture}[scale=.45, thick]
	
	\node[left] at (.5,0) {\tiny $S$=0};
	\node[left] at (.5,2) {\tiny $S$=1};
	\node[left] at (.5,4) {\tiny $S$=2};
	\node[left] at (.5,6) {\tiny $S$=3};
	
	\draw [very thin,gray,dashed] (.5,0) -- (18,0);
	\draw [very thin,gray,dashed] (.5,2) -- (18,2);
	\draw [very thin,gray,dashed] (.5,4) -- (18,4);
	\draw [very thin,gray,dashed] (.5,6) -- (18,6);

	\foreach \x in {1,3,5,7,9,11,13,15,17}{
		\draw [lightgray] (\x,0) -- (\x+.5,2);
	}
	\foreach \x in {2,4,6,8,10,12,14,16,18}{
	\draw [lightgray] (\x,0) -- (\x-.5,2);
	}

	\foreach \x in {1.5,7.5,13.5}{
	\draw [lightgray]  (\x,2) -- (\x+2,4);}
	\foreach \x in {3.5,9.5,15.5}{
	\draw [lightgray]  (\x,2) -- (\x,4);}
	\foreach \x in {5.5,11.5,17.5}{
	\draw [lightgray] (\x,2) -- (\x-2,4);}
	\foreach \x in {3.5}{
	\draw (\x,4) -- (\x+6,6);}
	\foreach \x in {9.5}{
	\draw (\x,4) -- (\x,6);}
	\foreach \x in {15.5}{
	\draw (\x,4) -- (\x-6,6);}

	\foreach \x in {1,2,3,4,5,7,13}{
	\fill [white] (\x,0) circle [radius=.15];
	\draw[very thin] (\x,0) circle [radius=.15];	
	}
	\foreach \x in {6,8,9,10,11,12,14,15,16,17,18}{
	\fill [gray] (\x,0) circle [radius=.15];
	\draw[very thin] (\x,0) circle [radius=.15];
	}
	
	\foreach \x in {1.5,3.5,5.5,7.5,9.5,11.5,13.5,15.5,17.5}{
	\fill [lightgray] (\x,2) circle [radius=.1];	
	}

	\fill [pink] (3.5,4) circle [radius=.15];
	\draw[very thin] (3.5,4) circle [radius=.15];
	\fill [green] (9.5,4) circle [radius=.15];
	\draw[very thin] (9.5,4) circle [radius=.15];
	\fill [green] (15.5,4) circle [radius=.15];
	\draw[very thin] (15.5,4) circle [radius=.15];

	\foreach \x in {9.5}{
	\fill [black] (\x,6) circle [radius=.1];
	}

	\end{tikzpicture}
		\caption{$\mathcal{PC}(18,11)$ with $\bm{G}=\bm{T_3} \otimes \bm{T_3} \otimes \bm{T_2}$}
		\label{fig:n18_fast_tree:b}
	\end{subfigure}
	}
	\caption{SC decoding trees (light grey) which are pruned to their Fast-SSC counterparts (black). White and grey leaf nodes represent frozen and information bits, respectively. Blue
	nodes are REP3A, pink nodes are REP3B, and green nodes are SPC.} 
	\label{fig:n18_fast_tree}
\end{figure*}

\subsection{Reduced Latency Nodes}
In this section, we will investigate the compatibility of the four basic reduced latency nodes with the ternary kernel. We will present generalizations
of each node type for any kernel ordering, though we also propose practical implementations that are valid under certain constraints. 

\subsubsection{Rate-0 Node}

A Rate-0 nodes one that leads to a set of leaf nodes with indices $i \in \mathcal{F}$ does not need to be traversed further since it is known 
that all $p$ leaf nodes and partial sums are 0 \cite{Alamdar-Yazdi2011}. Further, any node that leads to a set of intermediate nodes that are 
all Rate-0 is also Rate-0. This relationship does not change when considering the ternary kernel since encoding $(0,0,0)$ with $\bm{T_3}$ results in $(0,0,0)$. 

\subsubsection{Rate-1 Node}

Any node that is parent to leaf nodes whose indices $i \in \mathcal{I}$ is considered a Rate-1 node, as are any nodes that are parent to only 
Rate-1 nodes. Rate-1 nodes can be decoded by setting the $p$ partial sums at that stage with a hard decision on each soft LLR in that node and 
then updating the values of the leaf nodes by hard decoding the partial sums \cite{Alamdar-Yazdi2011}. Specifically, the partial sums vector $\bm{\beta_v}$ 
in node $v$ are obtained from the LLR vector $\bm{\alpha_v}$ by $\bm{\beta_v} = h(\bm{\alpha_v})$. The estimated sourceword values in 
$\bm{\hat{u}_v} \subset \bm{\hat{u}}$ with indices in $\mathcal{I}_v \subset \mathcal{I}$ are hard decoded using $\bm{\hat{u}_v} = \bm{\beta_v}\bm{G}_{p}^{-1}$ 
where $\bm{G}_{p}$ is the generator matrix obtained by performing the Kronecker product of kernels pertaining to all stages below that of node $v$.
Alternatively, $\bm{\hat{u}_v}$ may be estimated by propagating the partial sums $\bm{\beta_v}$ to the leaf nodes with inverse partial sum equations 
for each stage where $c_2^{-1} = c_2$ and $c_3^{-1}(s_0, s_1, s_2) = (s_0 \oplus s_1 \oplus s_2, s_1 \oplus s_2, s_0 \oplus s_2)$.

The following is a proof that outlines the validity of this decoding method for stages corresponding to $\bm{T_3}$.
Each Rate-1 node $v$ has the property that
\begin{equation}
	\label{eq:rate_1}
	\bm{\beta_{v_l}} = h(\bm{\alpha_{v_l}}), \bm{\beta_{v_c}} = h(\bm{\alpha_{v_c}}), \bm{\beta_{v_r}} = h(\bm{\alpha_{v_r}}).
\end{equation}
where $\bm{\alpha_{v_l}}$, $\bm{\alpha_{v_c}}$, and $\bm{\alpha_{v_r}}$ are the LLRs in the three branches below node $v$, as depicted in Fig. \ref{fig:n18_fast_tree:a}.
For shorthand, let $e = 3i$, $o = 3i + 1$, and $u = 3i + 2$ for all fixed $i$. Recall that the $\boxplus$ operator has the property
\begin{equation}
	h(a \boxplus b) = h(a) \oplus h(b) \text{ if } ab \ne 0.
\end{equation}
And so assuming that $\alpha_v[e] \ne 0$, $\alpha_v[o] \ne 0$, and $\alpha_v[u] \ne 0$ indicates that 

\small
\vspace*{-\baselineskip}
\begin{align}
	h(\alpha_{v_l}[i]) &= h(\alpha_v[e]) \oplus h(\alpha_v[o]) \oplus h(\alpha_v[u]) ; \text{ thus} \nonumber \\
	h(\alpha_{v_c}[i]) & \stackrel{(a)}{=} h(\alpha_v[o] \boxplus  \alpha_v[u]  + (1-2h(\alpha_{v_l}[i]))\alpha_v[e]) \nonumber \\
						&= h(\alpha_v[o] \boxplus  \alpha_v[u]  \nonumber  \\ 
						& \qquad + (1-2(h(\alpha_v[e]) \oplus h(\alpha_v[o]) \oplus h(\alpha_v[u])))\alpha_v[e]) \nonumber \\
						&= h(\alpha_v[o] \boxplus  \alpha_v[u]) = h(\alpha_v[o]) \oplus  h(\alpha_v[u]) \nonumber \\
	h(\alpha_{v_r}[i]) & \stackrel{(a)}{=} h((1\!-\!2h(\alpha_{v_l}[i]))\alpha_v[o] \nonumber \\
						& \qquad + (1\!-\!2h(\alpha_{v_l}[i] \!\oplus\! \alpha_{v_c}[i]))\alpha_v[u]) \nonumber \\
						&= h((1-2(h(\alpha_v[e]) \oplus h(\alpha_v[o]) \oplus h(\alpha_v[u])))\alpha_v[o] \nonumber \\
						& \qquad + (1-2(h(\alpha_v[e]) \oplus h(\alpha_v[o]) \oplus h(\alpha_v[u]) \nonumber \\
						& \qquad \qquad \oplus h(\alpha_v[o]) \oplus h(\alpha_v[u]) ))\alpha_v[u]) \nonumber \\
						&= h(\alpha_v[e]) \oplus  h(\alpha_v[u]) \nonumber
\end{align}
\vspace*{-.2\baselineskip}
\normalsize
where (a) uses eq. \ref{eq:rate_1}. Further, 
\vspace*{-.1\baselineskip}
\small
\begin{align*}
	\beta_v[u] &= \beta_{v_l}[i] \oplus \beta_{v_c}[i] \oplus \beta_{v_r}[i]  \\
			& \stackrel{(b)}{=} h(\alpha_{v_l}[i]) \oplus h(\alpha_{v_c}[i]) \oplus h(\alpha_{v_r}[i])   \\
			&= h(\alpha_v[e]) \oplus h(\alpha_v[o]) \oplus h(\alpha_v[u])  \\
			& \qquad \oplus h(\alpha_v[o]) \oplus  h(\alpha_v[u])  \\
			& \qquad \oplus h(\alpha_v[e]) \oplus  h(\alpha_v[u]) = h(\alpha_v[u]); \text{ thus}  \\
	\beta_v[o] &= \beta_{v_l}[i] \oplus \beta_{v_r}[i] \stackrel{(b)}{=} h(\alpha_{v_l}[i]) \oplus h(\alpha_{v_r}[i])  \\
			   &= h(\alpha_v[o])  \\
	\beta_v[e] &= \beta_{v_l}[i] \oplus \beta_{v_c}[i] \stackrel{(b)}{=} h(\alpha_{v_l}[i]) \oplus h(\alpha_{v_c}[i])  \\
			   &= h(\alpha_v[e]) 
\end{align*}
\normalsize
where (b) also makes use of eq. \ref{eq:rate_1}. Hence, the proof is completed and asserts confirmation that decoding of a Rate-1 node can be computed using the same method
for $\bm{T_3}$ as for $\bm{T_2}$. The original proof in \cite{Alamdar-Yazdi2011} describes through induction that this method of decoding holds for a Rate-1 node of
any depth.  

\subsubsection{Single-Parity Check Node}
For an Ar{\i}kan polar code of rate $R = \frac{N-1}{N}$, the lowest order bit $u_0$ will be frozen. Such a polar code can be interpreted as
a single-parity check (SPC) code. For instance, if $N=4$, then $\bm{u}=(0,a_0,a_1,a_2)$ and $\bm{x}=(a_0 \oplus a_1 \oplus a_2, a_0\oplus a_2, a_1 \oplus a_2, a_2)$. 
An SC node $v$ with $\mathcal{F}_v$ that reflects this pattern can be optimally decoded with low complexity. First, the partial sums $\bm{\beta_v}$ are estimated
with a hard decision in the usual fashion, \emph{ie} $\bm{\beta_v} = h(\bm{\alpha_v})$. The parity of $\bm{\beta_v}$ is then computed using
\begin{equation}
	\text{parity } = \bigoplus_{i=0}^{N_v - 1} h(\beta_v[i]).
\end{equation}
If the parity constraint is not fulfilled, then the least reliable bit at index $j$ is flipped as in 
\begin{equation}
	\beta_v[j] := \beta_v[j] \oplus \text{parity},
\end{equation}
where $j = \argmin_i|\alpha_v[i]|$. Finally, $\bm{\hat{u}_v}$ is computed using the same hard decoding procedure as a Rate-1 node. 

In MK polar codes with rate $R = \frac{N-1}{N}$, $u_0$ will always be frozen regardless of the order of kernels, and so MK polar codes also have an embedded SPC property. 
Observe that if $N=3$, then $\bm{u}=(0,u_0,u_1)$ and $\bm{x}=(u_0, u_1, u_0 \oplus u_1)$. As such, SPC nodes can be identified and decoded in exactly the same way as with 
polar codes using only the Ar{\i}kan kernel. 

\subsubsection{Repetition Node}
\label{section:fssc:rep}

In an Ar{\i}kan polar code with rate $R = \frac{1}{N}$, only the highest order bit $u_{N-1}$ will contain information. This frozen set pattern renders
the polar code into a repetition (REP) code. Any node $v$ identified as a REP node is decoded simply by summing all soft LLRs and taking a hard 
decision on the result. The resulting bit is stored in all indices of $\bm{\beta_v}$:
\begin{equation}
		\label{eq:rep_sum}
	\beta_v[i] = h\bigg(\sum_j \alpha_v[j]\bigg) \text{ for } i,j \in v
\end{equation}
The partial sum $\bm{\beta_v}$ can then be hard decoded to compute $\bm{\hat{u}_v}$. Alternatively, $\bm{\hat{u}_v}$ can be evaluated more efficiently with
\begin{equation}
	\label{eq:rep_u}
	\bm{\hat{u}_v}[i] = \begin{cases} 0 & i < N-1 \\ h(\sum_j \alpha_v[j]) & i = N-1 \end{cases} \text{ for } i,j \in v.
\end{equation}
The ternary kernel also mirrors repetition codes when presented with the same frozen set pattern, although the decoding procedures are more involved. Unlike the 
Ar{\i}kan kernel, when the highest order bit of a ternary kernel is the only information bit, the data is not repeated in all codeword bits. Specifically, if $N=3$ and
$\bm{u}=(0,0,a_0)$, then $\bm{x}=(0, a_0, a_0)$. In this example, repetition decoding can still be carried out, supposing the first index is excluded in the sum in eq. \ref{eq:rep_sum}. 
Generally, a REP node $v$ at stage $S$ has REP pattern $P_v$, which is determined for any combination of Ar{\i}kan or ternary kernels by {}
performing a recursive Kronecker product of repetition patterns $P_2 = (1,1)$ or $P_3=(0,1,1)$:

\vspace*{-\baselineskip}
\begin{equation}
	P_v = \bigotimes_{i=0}^{S-1} P_{k_i}.
\end{equation}

Eq. \ref{eq:rep_sum} can then be modified to accommodate this new constraint: 

\vspace*{-.5\baselineskip}
\begin{equation}
		\label{eq:rep_sum_mk}
	\beta_v[i] = h\bigg(\sum_j \alpha_v[j] \cdot P_v[j]\bigg) \text{ for } i,j \in v
\end{equation}
Table \ref{table:rep_patterns} outlines several examples of REP patterns with varying kernel orders. Repetition nodes can be appropriately labeled for decoder scheduling purposes depending on 
the order of kernels. A repetition node that is made up of only Ar{\i}kan kernels, previously the only type of REP node, can now be labeled as a \textit{REP2} node. REP
nodes comprised of only ternary kernels can be labeled as \textit{REP3A} nodes. Mixed kernel repetition nodes can be labeled \textit{REP3B} or \textit{REP3C} depending 
on the order of $\bm{T_2}$ or $\bm{T_3}$.

Although a MK polar code can be built using any arbitrary order of kernels, it often is the case that the non-Ar{\i}kan kernels are either the first or last constituents in
the Kronecker expression used to compute $\bm{G}$. Further, Section \ref{section:kernel_order} demonstrates that assuming the order of kernels without optimization presents
comparable error correction results. Moreover, using LDPC WiMAX code lengths as a guideline \cite{Shin2012} suggests that MK polar codes are able to achieve most desired code lengths
with only a few stages of non-Ar{\i}kan kernels. Considering this behaviour of kernel orders in MK codes, it is unnecessary to implement generalized
ternary repetition nodes as the majority of cases can be efficiently decoded under a few constraints. As such, the scheduling and implementation of MK REP nodes can be simplified by
eliminating $P_v$ computation. We limit REP3A nodes to have a maximum size of 27 so that there are only 3 possible $P_v$ that
can simply be stored instead of computed. Additionally, we limit REP3B and REP3C nodes to contain only a single ternary stage so that computation of $\bm{\beta_v}$ can be 
carried out efficiently: 

\vspace*{-\baselineskip}
\begin{align}
	\beta_{v[i]_{\text{REP3B}}} &= h\bigg(\sum_j \alpha_v[j]\bigg) \text{ for } i, j \in [\frac{N_v}{3}, N_v), \nonumber \\
	\beta_{v[i]_{\text{REP3C}}} &= h\bigg(\sum_j \alpha_v[j]\bigg) \text{ for } i \text{ and } j \not\equiv \text{ mod }3. \nonumber
\end{align}

The summation in eq. \ref{eq:rep_sum} can be modified for REP3B nodes by skipping the first third of indices, while for REP3C node it is modified
by skipping every third index. Of course, eq. \ref{eq:rep_u} still applies to ternary repetition nodes. Under the requirement of only a single ternary stage,
these nodes do not need to be limited in size. We will utilize these simplifications in our numerical analysis.

\begin{table}
	\centering
	\begin{tabular}{@{}llcr@{}}
	\toprule
	$N_v$ & $\bm{k_v}$ & $P_v$                                 & Type  \\
	\cmidrule(r){1-1}
	\cmidrule(l){2-2}
	\cmidrule(l){3-3}
	\cmidrule(l){4-4}
	3     & (3)        & (0,1,1)                               & REP3A \\
	6     & (2,3)      & (0,1,1,0,1,1)                         & REP3C \\
	6     & (3,2)      & (0,0,1,1,1,1)                         & REP3B \\
	8     & (2,2,2)    & (1,1,1,1,1,1,1,1)                     & REP2  \\
	9     & (3,3)      & (0,0,0,0,1,1,0,1,1)                   & REP3A \\
	12    & (2,2,3)    & (0,1,1,0,1,1,0,1,1,0,1,1)             & REP3C \\
	12    & (3,2,2)    & (0,0,0,0,1,1,1,1,1,1,1,1)             & REP3B \\
	18    & (2,3,3)    & (0,0,0,0,1,1,0,1,1,0,0,0,0,1,1,0,1,1) & REP3C \\
	\bottomrule
	\end{tabular}
		\caption{Examples of $P_v$ patterns.}
			\vspace*{-\baselineskip}
	\label{table:rep_patterns}
\end{table}

\section{Complexity Reduction Evaluation}
\label{section:analysis}

\begin{table*}[t]
	\footnotesize
	\centering
	\begin{tabular}{@{}lrrrrrrrrr@{}}
	\toprule
	$N$                   & $R$  & \# SC Nodes & \# Fast-SSC Nodes & \# R0 & \# R1 & \# SPC & \# REP2 & \# REP3A/B/C & \% Reduction \\
	\cmidrule(r){1-1}\cmidrule(r){2-2}\cmidrule(r){3-3}\cmidrule(r){4-4}\cmidrule(r){5-5}\cmidrule(r){6-6}\cmidrule(r){7-7}\cmidrule(r){8-8}\cmidrule(r){9-9}\cmidrule(l){10-10}
	\multirow{3}{*}{96}   & 0.25 & 158/189     & 37/27             & 7/2   & 1/0   & 4/4    & 0/4     & 1/0     & 76.6/85.7    \\
	                      & 0.5  & 158/189     & 43/45             & 8/5   & 1/5   & 6/3    & 0/3     & 0/0     & 72.8/76.2    \\
	                      & 0.75 & 158/189     & 37/42             & 3/3   & 5/6   & 4/4    & 0/3     & 0/1     & 76.6/77.8    \\
	\cmidrule(r){1-1}\cmidrule(r){2-2}\cmidrule(r){3-3}\cmidrule(r){4-4}\cmidrule(r){5-5}\cmidrule(r){6-6}\cmidrule(r){7-7}\cmidrule(r){8-8}\cmidrule(r){9-9}\cmidrule(l){10-10}
	\multirow{3}{*}{432}  & 0.25 & 654/849     & 101/118           & 15/11 & 4/6   & 16/13  & 0/11    & 4/2     & 84.5/86.1    \\
	                      & 0.5  & 654/849     & 110/136           & 14/9  & 4/7   & 21/19  & 0/15    & 7/0     & 83.2/83.4    \\
	                      & 0.75 & 654/849     & 106/109           & 13/9  & 9/9   & 17/14  & 0/8     & 2/0     & 83.8/87.2    \\
	\cmidrule(r){1-1}\cmidrule(r){2-2}\cmidrule(r){3-3}\cmidrule(r){4-4}\cmidrule(r){5-5}\cmidrule(r){6-6}\cmidrule(r){7-7}\cmidrule(r){8-8}\cmidrule(r){9-9}\cmidrule(l){10-10}
	\multirow{3}{*}{768}  & 0.25 & 1278/1533   & 196/186           & 34/17 & 5/8   & 24/19  & 0/19    & 3/0     & 84.6/87.9    \\
	                      & 0.5  & 1278/1533   & 223/222           & 31/15 & 9/14  & 31/24  & 0/22    & 4/0     & 82.5/85.5    \\
	                      & 0.75 & 1278/1533   & 172/192           & 19/12 & 10/19 & 25/19  & 0/15    & 4/0     & 86.5/87.5    \\
	\cmidrule(r){1-1}\cmidrule(r){2-2}\cmidrule(r){3-3}\cmidrule(r){4-4}\cmidrule(r){5-5}\cmidrule(r){6-6}\cmidrule(r){7-7}\cmidrule(r){8-8}\cmidrule(r){9-9}\cmidrule(l){10-10}
	\multirow{3}{*}{2304} & 0.25 & 3582/4602   & 409/453           & 62/31 & 8/16  & 71/54  & 0/52    & 5/0     & 88.6/90.1    \\
	                      & 0.5  & 3582/4602   & 487/516           & 63/23 & 17/17 & 86/78  & 0/56    & 8/0     & 86.4/88.8    \\
	                      & 0.75 & 3582/4602   & 395/441           & 45/24 & 27/39 & 60/50  & 0/36    & 9/0     & 88.9/90.4    \\
	\bottomrule
	\end{tabular}
	\caption{\small Latency reduction of MK polar codes with ternary kernels as last/first Kronecker constituents.}
	\label{table:latency_reduction}
\end{table*}

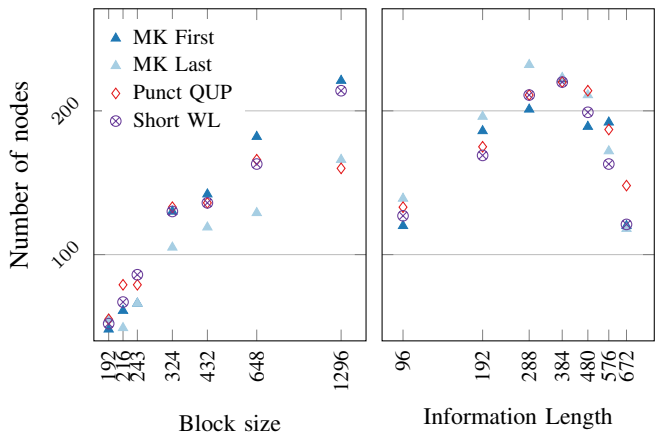
\begin{figure}[t]
\captionsetup[subfigure]{position=t}
\makebox[\linewidth][c]{
	\hspace{-5mm}
	\centering
		\subcaptionbox*{}{
			\begin{tikzpicture}
\begin{semilogxaxis}[height=6cm,width=5.2cm,
        scaled ticks=false,
        enlargelimits={upper=0.3},
        ymin=40,ymax=250,
        xmin=170,xmax=1298,
        log basis x =2,
        xtick={192,216,243,324,432,648,1296},
        xticklabels={192,216,243,324,432,~~648,1296},
        xticklabel style={rotate=90,  font=\footnotesize},
        yticklabel style={rotate=45, font=\scriptsize},
        xlabel={Block size},
        xlabel style={font=\small},
        ylabel={Number of nodes},
        ymajorgrids,
        legend entries={~MK First, ~MK Last, ~Punct QUP, ~Short WL},  
        legend style={font=\small, draw=none, nodes={scale=0.9}},
        legend pos=north west,
        legend cell align={left},
    ]

        \addplot[Paired-1, only marks, mark=triangle*] table [x={N}, y={FSSC}] {figs/data/k164/r_mk_l_2048.txt};
        \addplot[Paired-2, only marks, mark=triangle*] table [x={N}, y={FSSC}] {figs/data/k164/r_mk_f_2048.txt};
        \addplot[Paired-5, only marks, mark=diamond] table [x={N}, y={FSSC}] {figs/data/k164/r_punct_first_2048.txt};
        \addplot[Paired-9, only marks, mark=otimes] table [x={N}, y={FSSC}] {figs/data/k164/r_short_last_2048.txt};

\end{semilogxaxis}  
\end{tikzpicture}
		}
		\hspace{-7.5mm}		
		\subcaptionbox*{}{
			\begin{tikzpicture}
\begin{semilogxaxis}[height=6cm,width=5.2cm, 
        scaled ticks=false,
        enlargelimits={upper=0.3},
        ymin=40, ymax=250,
        xmin=80,
        log basis x =2,
        xtick={96 ,192 ,288 ,384 ,480 ,576 , 672},
        xticklabels={96 ,192 ,288 ,384 ,~480 ,576 ,672},
        xticklabel style={rotate=90, font=\footnotesize},
        yticklabel style={rotate=45, font=\scriptsize, draw=none},
        yticklabels={,,},
        xlabel={Information Length},
        xlabel style={font=\small},
        ymajorgrids,
        xmin=80,xmax=690,
        ymin=40,ymax=250,
    ]

        \addplot[Paired-1, only marks, mark=triangle*] table [x={K}, y={FSSC}] {figs/data/n768/r_mk_l_2048.txt};
        \addplot[Paired-2, only marks, mark=triangle*] table [x={K}, y={FSSC}] {figs/data/n768/r_mk_f_2048.txt};
        \addplot[Paired-5, only marks, mark=diamond] table [x={K}, y={FSSC}] {figs/data/n768/r_punct_first_2048.txt};
        \addplot[Paired-9, only marks, mark=otimes] table [x={K}, y={FSSC}] {figs/data/n768/r_short_last_2048.txt};

\end{semilogxaxis}  
\end{tikzpicture}
		}
	}
	\caption{Fast-SSC complexity for (a) $K=164$ and (b) $N=768$ sweeping rates $R\approx\frac{1}{8},\dots,\frac{7}{8}$.}
	\label{fig:fssc_plot}
\end{figure}

This section outlines the effectiveness of MK compatible Fast-SSC with numerical examples. With a sufficiently large processor, it can be assumed that each node in a decoding 
tree constitutes a single operation. As such, all measurements of decoding complexity refer to the number of decoding nodes. All polar codes analyzed were constructed using GA 
with a target $Eb/N0$ of 3dB adopting a BPSK modulation scheme, \emph{ie} $\mathcal{M}=1$. Fig. \ref{fig:fssc_plot} compares the number of computations in Fast-SSC decoders for various 
length-compatible polar codes over a range of codeword lengths and rates. \textit{Punct QUP} \cite{Niu2013} and \textit{Short WL} \cite{Wang2014} indicate puncturing and 
shortening patterns that allow for large numbers of frozen sets to be grouped together. Generally, \textit{Punct QUP} and \textit{Short WL} methods have comparable complexity 
to both kernel orderings of MK polar codes. However, MK codes with a high proportion of ternary stages, such as lengths $N=(216, 324, 648)$, have the fewest decoding operations
overall when built with the \textit{Last} kernel ordering. This is a result of the fast decoders for ternary nodes, which decode a larger number of bits simultaneously than would Ar{\i}kan
nodes. This is depicted in Fig. \ref{fig:n18_fast_tree:a} where a node in stage $S=2$ decodes 9 bits at once, where as node in the same stage of Fig. \ref{fig:n18_fast_tree:b} decodes only 6 bits 
at once. Therefore, it may be desirable to construct MK polar codes using the \textit{Last} kernel ordering from a decoding complexity standpoint for particular code lengths.

Regarding latency reduction of MK decoding, Table \ref{table:latency_reduction} outlines various comparisons between SC and Fast-SSC along with the node makeup of Fast-SSC 
decoding schedules. Just as with Ar{\i}kan polar codes, there is a greater latency reduction for extreme rates. This gain is due to the higher proportion of Rate-1 and SPC nodes for
high rates and Rate-0 and REP nodes for low rates. Observe that longer MK polar codes have greater latency reduction than short polar codes. It should also be noted 
that MK polar codes with a higher number of ternary stages have increased latency with the \textit{First} kernel permutation compared with \textit{Last}. Further, the number of 
ternary repetition nodes is proportionately low, so it is sufficient to consider only the most common cases for a simplified implementation rather
than a generalized algorithm as outlined in Section \ref{section:fssc:rep}. 

\section{Conclusion}
\label{section:conclusion}

In this work, we extended the Fast-SSC polar code decoder to be compatible with MK polar codes.
The four basic fast nodes were generalized to be able to decode MK codes with arbitrary kernel orders.
We propose simplified implementations of ternary repetition nodes that are valid under imposed constraints that reflect
typical MK behavior. We have observed that in all tested cases, a minimum 72\% latency reduction of SC is 
achievable for MK polar codes. This study can act as a guideline for future work on a hardware implementation of a fast
MK decoder.

\bibliographystyle{IEEEtran}
\bibliography{IEEEabrv,mk_fast}

\end{document}